\documentclass[reprint, aps, showpacs]{revtex4-1}
\usepackage[utf8]{inputenc}
\usepackage{graphicx}
\usepackage{amsmath}
\usepackage{tikz}
\usepackage{xcolor}
\usepackage{upgreek}
\usepackage{ulem}

\begin{document}

\title{Spatial profile of accelerated electrons from ponderomotive scattering in hydrogen cluster targets}

\author{B. Aurand}
\email{bastian.aurand@hhu.de}
\affiliation{Institut f{\"u}r Laser- und Plasmaphysik, Heinrich-Heine-Universit{\"a}t D\"{u}sseldorf, 40225 D{\"u}sseldorf, Germany}

\author{L. Reichwein}
\email{lars.reichwein@hhu.de}
\affiliation{Institut f{\"u}r Theoretische Physik I, Heinrich-Heine-Universit{\"a}t D\"{u}sseldorf, 40225 D{\"u}sseldorf, Germany}

\author{K. M. Schwind}
\affiliation{Institut f{\"u}r Laser- und Plasmaphysik, Heinrich-Heine-Universit{\"a}t D\"{u}sseldorf, 40225 D{\"u}sseldorf, Germany}

\author{E. Aktan}
\affiliation{Institut f{\"u}r Laser- und Plasmaphysik, Heinrich-Heine-Universit{\"a}t D\"{u}sseldorf, 40225 D{\"u}sseldorf, Germany}

\author{M. Cerchez}
\affiliation{Institut f{\"u}r Laser- und Plasmaphysik, Heinrich-Heine-Universit{\"a}t D\"{u}sseldorf, 40225 D{\"u}sseldorf, Germany}

\author{V. Kaymak}
\affiliation{Institut f{\"u}r Theoretische Physik I, Heinrich-Heine-Universit{\"a}t D\"{u}sseldorf, 40225 D{\"u}sseldorf, Germany}

\author{L. Lessmann}
\affiliation{Institut f{\"u}r Kernphysik, Westf{\"a}lische Wilhelms-Universit{\"a}t M\"{u}nster, 48149 M{\"u}nster, Germany}

\author{R. Prasad}
\affiliation{Institut f{\"u}r Laser- und Plasmaphysik, Heinrich-Heine-Universit{\"a}t D\"{u}sseldorf, 40225 D{\"u}sseldorf, Germany}

\author{J. Thomas}
\affiliation{Institut f{\"u}r Theoretische Physik I, Heinrich-Heine-Universit{\"a}t D\"{u}sseldorf, 40225 D{\"u}sseldorf, Germany}

\author{T. Toncian}
\affiliation{Institut f{\"u}r Strahlenphysik, Helmholtz-Zentrum Dresden-Rossendorf, 01328 Dresden, Germany}

\author{A. Khoukaz}
\affiliation{Institut f{\"u}r Kernphysik, Westf{\"a}lische Wilhelms-Universit{\"a}t M\"{u}nster, 48149 M{\"u}nster, Germany}

\author{A. Pukhov}
\affiliation{Institut f{\"u}r Theoretische Physik I, Heinrich-Heine-Universit{\"a}t D\"{u}sseldorf, 40225 D{\"u}sseldorf, Germany}

\author{O. Willi}
\affiliation{Institut f{\"u}r Laser- und Plasmaphysik, Heinrich-Heine-Universit{\"a}t D\"{u}sseldorf, 40225 D{\"u}sseldorf, Germany}

\date{\today}

\begin{abstract}
	We study the laser-driven acceleration of electrons from overdense hydrogen clusters to energies of up to 13\,MeV in laser forward direction and several hundreds of keV in an outer ring-like structure. The use of cryogenic hydrogen allows for high repetition-rate operation and examination of the influence of source parameters like temperature and gas flow. The outer ring-like structure of accelerated electrons, originating from the interaction, that is robust against the change of laser and target parameters can be observed for low electron densities of ca. 3$\times$10$^{16}$\,cm$^{-3}$.  For higher electron densities, an additional central spot of electrons in the laser forward direction can be observed. Utilizing 3D-PIC simulations, it is revealed that both electron populations mainly stem from ponderomotive scattering.
\end{abstract}

\maketitle
\section{Introduction}
Clustered targets and their interaction with high-intensity laser pulses have been studied for a long time, i.a. due to the possibility for high repetition-rate operation when accelerating particles \cite{Cang2004, Aurand2019, Aurand2020} and increased x-ray emission \cite{Lecz2020}.
In general, laser-plasma based acceleration schemes have received great interest because of the high achievable electromagnetic field strengths compared to conventional accelerators \cite{Padamsee2017}. One of the most prominent examples is the technique of laser-driven plasma wakefield acceleration (LWFA), for which a short, highly intense laser pulse is interacting with a plasma and pushes electrons away from the region of the highest intensity due to the ponderomotive force, leaving behind an electronic cavity \cite{Tajima1979, Pukhov2002}. Choosing appropriate laser and plasma parameters, a so-called bubble can be excited. This bubble is an almost spherical structure, traversing the plasma with velocities close to the speed of light $c$ that exhibits uniform accelerating fields \cite{Faure2004}.
As soon as a self-trapping condition is met, electrons can be trapped and accelerated to high energies. Over the last years, several more advanced setups have been proposed that aim to improve on some aspects of LWFA like beam emittance \cite{Hidding2012} or increasing the injection of electrons into the wake \cite{Oz2007, Ekerfelt2017}.	As there is a further trend going to higher-density targets \cite{Wettervik2018} due to the stronger fields that can be applied, cluster targets and targets doped with nano-particles have been considered in several publications \cite{Cho2018, Aniculaesei2019, Mayr2020, Lecz2020}.
Specifically in the works by Cho \textit{et al} \cite{Cho2018} and Aniculaesei \textit{et al} \cite{Aniculaesei2019}, the presence of nano-particles was shown to aid electron injection in LWFA similar to ionisation injection \cite{Oz2007}.
Mayr \textit{et al} \cite{Mayr2020} recently have numerically considered a target consisting of Argon clusters which has shown that the injection mechanism in the presence of clusters is modified: electrons are ripped from the clusters by the laser pulse and can be injected if they are close to the central axis. This allows for a continuous injection on electrons due to the random positions of clusters within the target volume.
	
Besides LWFA there are several other acceleration processes that can become important when considering clustered targets. Lower energy electrons may be ejected via the ponderomotive potential of the laser pulse interaction with the target, a process referred to as ponderomotive scattering \cite{Hartemann1995}. This mechanism can be viewed as a one-time momentum transfer that accelerates the electrons. In contrast, in direct laser acceleration (DLA), the electrons get accelerated by the laser pulse over a longer period of time \cite{Jirka2020}. This generally yields higher energies than ponderomotive scattering, but the process is strongly dependent on the plasma target parameters as the electrons need to stay in phase with the laser.
	
In this paper, the acceleration of electrons is considered for a hydrogen-cluster target that has so far been used for the study on the generation of laser-driven proton beams by Coulomb-explosion \cite{Aurand2019, Aurand2020}.  After a short overview of the experimental setup, the experimental results obtained from different types of diagnostics are presented. The resulting spatial structure of accelerated electrons consists of a central spot in laser forward direction and a surrounding outer ring, similar to structures observed from LWFA \cite{Brunetti2017, Behm2019, Pollock2015}. There, the outer ring stems from electrons that do not meet the trapping condition in the first wake but get accelerated via the subsequent wakes. For the lower densities used in the studies presented here, however, the main mechanism of electron acceleration is identified to be ponderomotive scattering by the focused laser pulse. As found by Hartemann \textit{et al} \cite{Hartemann1995}, low-energy electrons are ejected with a larger emission angle than those with higher energies. This ring structure is found to be very stable with respect to changes in the laser-plasma parameters. These two spatial features are discussed in the scope of already existing analytic theories for ponderomotive scattering \cite{Hartemann1995, Quesnel1998} and 3D-PIC simulations. Further, the influence of the prepulse on the cluster target is discussed. Its presence leads to a Coulomb explosion before the main pulse hits and creates a more homogeneous plasma target. Depending on the density of this homogeneous background as well as the remaining clusters, more electrons get accelerated via LWFA.

\section{Setup}\label{sec:setup}
\begin{figure}[ht]
	\centering	
	\includegraphics[width=\columnwidth]{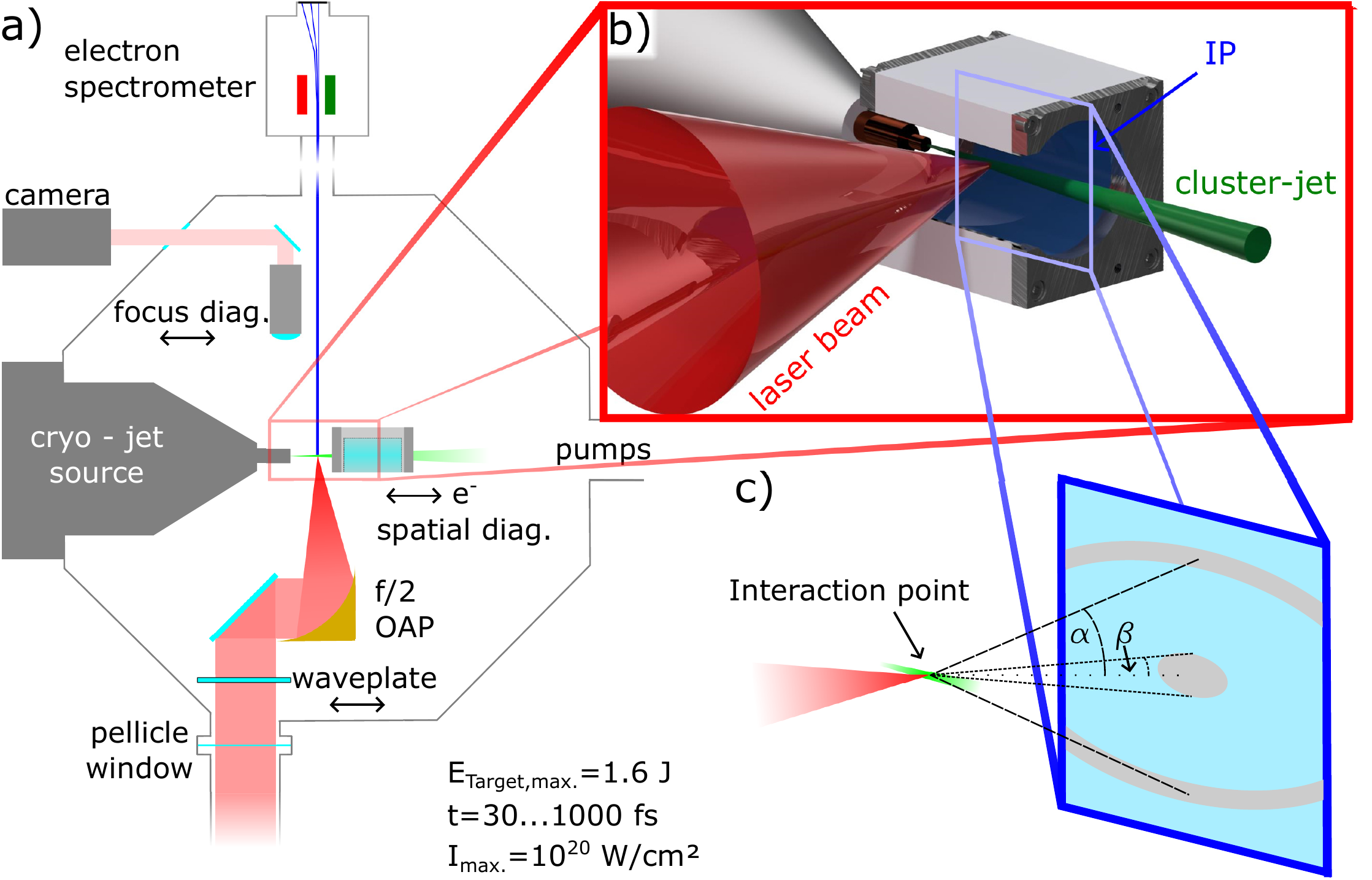}
	\caption{\label{fig1} a) Setup of the experiment. The laser beam was focused by an f/2-OAP into the centre of the cluster jet and the accelerated electrons were either recorded by a energy-resolving spectrometer with a MCP as a detector in the $0^{\circ}$ direction or by an IP mounted cylindrical around the interaction point. b) Close-up sketch of the IP geometry. c) Definition of the opening angles $\alpha$ and $\beta$ for the observed structures on the IP (see fig. \ref{fig3} and \ref{fig4}).}
\end{figure}

The experiment took place at the ARCTURUS laser facility, a 200 TW Ti:sapphire based laser-system with a double-CPA based amplification structure at the University of D\"{u}sseldorf \cite{Cerchez2019}.
Pulse energies of up to 7 J before compression at a minimal pulse duration of about 30 fs in 5 Hz operation utilizes an on-target intensity of $10^{20}$ W/cm$^2$. The experimental setup is shown in fig. \ref{fig1}a). In front of the target chamber a 1\,mm-thick anti-reflex coated pellicle-window is used to decouple the vacuum-systems of target chamber and laser-system to prevent gas or other hydro-carbon contamination in the optical compressor. Entering the experimental chamber, the laser is p-polarised with respect to the breadboard but the polarisation can be rotated to s- or changed into circular (c-polarised) by means of a half wave-plate or a quarter wave-plate, respectively.
The initial laser-beam is focused by an f/2-off-axis-parabola into the center of a hydrogen-cluster beam. At the interaction position a focal spot diameter of $w_0 = (5 \pm 0.4)$ $\upmu$m and a Rayleigh length of about $z_R = 30$ $\upmu$m is reached. Cross-calibrating the laser energy at the end of the amplifier chain with the laser energy on target resulted in a throughput of $(31 \pm 2)\%$ for the compressor and the beamline. 

The clusters are formed by a continuously operating cryogenic-cluster target, developed for the investigation of laser-plasma interactions, basically described in the paper by Grieser \textit{et al} \cite{Grieser2019}. For the study presented here, hydrogen was used as the target material. The gas is pre-cooled by a closed-loop helium cold-head to temperatures as low as $T = 24$ K and afterwards expanded adiabatically by a backing pressure of up to $p = 16$ bar through a de-Laval nozzle.

This expansion, causing a further cool-down of the gas, allows the formation of condensation centers by three-body interactions once the van-der-Waals binding energy overcomes the relative kinetic energy between the molecules. More molecules can get bound, forming macroscopic clusters of up to several million molecules in each cluster. Note, that for temperatures $< 33.2$ K the hydrogen is in the liquid aggregation state before expanded through the nozzle, while above this temperature it is in the gaseous aggregation state.
The formation process from the gaseous phase can be characterized by the empirical \textit{Hagena parameter}  $\Gamma$ \cite{Hagena1987}, depending on the stagnation temperature -- which is temperature of the gas before exiting the nozzle -- and pressure as well as  nozzle geometry and gas parameters.

For the hydrogen target used in our experiments, \cite{Hagena1987} and \cite{Hagena1992} imply, that the number $N$ of atoms/molecules per clusters is given by

\begin{align}
	\Gamma (p, T) &= \frac{k \cdot p \text{ [mbar]} \cdot \left(\frac{0.74 \cdot d_N \text{ [$\upmu$m]}}{\tan \alpha_{1/2}}\right)^{0.85}}{T^{2.29} \text{ [K]}} \; , \\
	N (p, T) &= A_N \cdot \left( \frac{\Gamma}{1000}\right)^{\gamma_N} \; , \label{eq:num_atoms}
\end{align}
where $k = 184$, $A_N(\Gamma > 1800) = 33 $ and $\gamma_N (\Gamma > 1800) = 2.35$.
For the given nozzle geometry the narrowest inner diameter $d_N = 42\,\upmu$m and the half opening angle of the divergent outlet area $\alpha_{1/2} = 3.5^\circ$ are used. Assuming the clusters to be spherical with the number of atoms calculated via equation (\ref{eq:num_atoms}), and with the density of liquid hydrogen $\rho_{\mathrm{liqu.,H}_2} = 76.3$ kg\,/\,m$^3$ the cluster diameter is calculated and verified by means of Mie scattering measurements to be in the range of $d_\mathrm{cl,Hagena} = 10-80$\,nm for the operation parameters used in \cite{Grieser2019}. Note that the values derived above correspond to the average cluster size, while the full-size distribution is described by a log-norm function \cite{Smith1964}.

The formation of the cluster jet was characterised by means of Schlieren imaging, revealing a flat density profile, which underlines the ballistic motion of the clusters exiting the cluster jet target. The atomic density at the interaction point, which is about 22 mm away from the nozzle was calculated, taking into account the atomic density of a single cluster of about $4 \times 10^{22}$ cm$^{-3}$, the measured gas flow of the target system and the opening angle of the cluster jet. A density of about $4 \times 10^{16}$ cm$^{-3}$ and $3 \times 10^{16}$ cm$^{-3}$ at a gas temperature of 34 K or 40 K, respectively was derived. After passing the interaction point, the cluster jet is dumped by a root-pump system, sufficient to keep the ambient pressure in the interaction chamber in the range of $10^{-1 \dots -3}$ mbar.

As one diagnostic for the electrons, a spectrometer in the laser forward direction is used. The electrons are dispersed in a magnetic field of $B$\,=\,0.18\,T and impact afterwards on a chevron-design micro-channel plate (MCP) being converted in a photon signal recorded by a CCD camera. This allows for a good energy resolution in the range of 6\,-\,40 \,MeV, and is suitable for the high-repetition rate capability of the target. As a second diagnostic, image plates (IPs; Fuji: BAS-TR) were used to detect the spatial profile of the electrons and read out by a scanner after irradiation \cite{Zuo1996, Tanaka2005}. The IPs were fixed using three motorised cylindrical mounts stacked on top of each other (Fig. \ref{fig1}b), allowing three individual recordings per pumpdown-cycle of the interaction chamber.
Each mount could be moved in order to be centered around the interaction area, allowing the cluster jet to pass along the central axis. The laser is coupled in by $90^\circ$ with respect to the central axis/cluster direction. The cylinders have a diameter of 50 mm covering, beside the slit, an angle of about $270^\circ$ in one direction and about $100^\circ$ opening angle in the plane of the breadboard. All IPs were wrapped in a 14 $\upmu$m aluminium foil to block the laser light as well as protons. It is known from our previous studies reported in \cite{Aurand2019, Aurand2020} that the proton energies achieved by the Coulomb-explosion are well below 1 MeV, which means that the Al-filter blocks them sufficiently.

Additionally, by adding a strong magnet into the cylindrical mount of the IPs for one irradiation run it was verified that the signal stems from electrons and not x-rays, because the magnetic-field caused a deflection of the observed structure, which underlined that the signal originated from charged particles, which can be electrons only. Fig \ref{fig1}c) shows a sketch of the signal observed on the IPs (cmp. fig \ref{fig3}). The opening angle enclosing the outer-ring structure is $\alpha$, while the opening angle of the central spot is $\beta$.

\section{Measurements}\label{sec:measurements}
\begin{figure}[ht] 
	\centering	
	\includegraphics[width=\columnwidth]{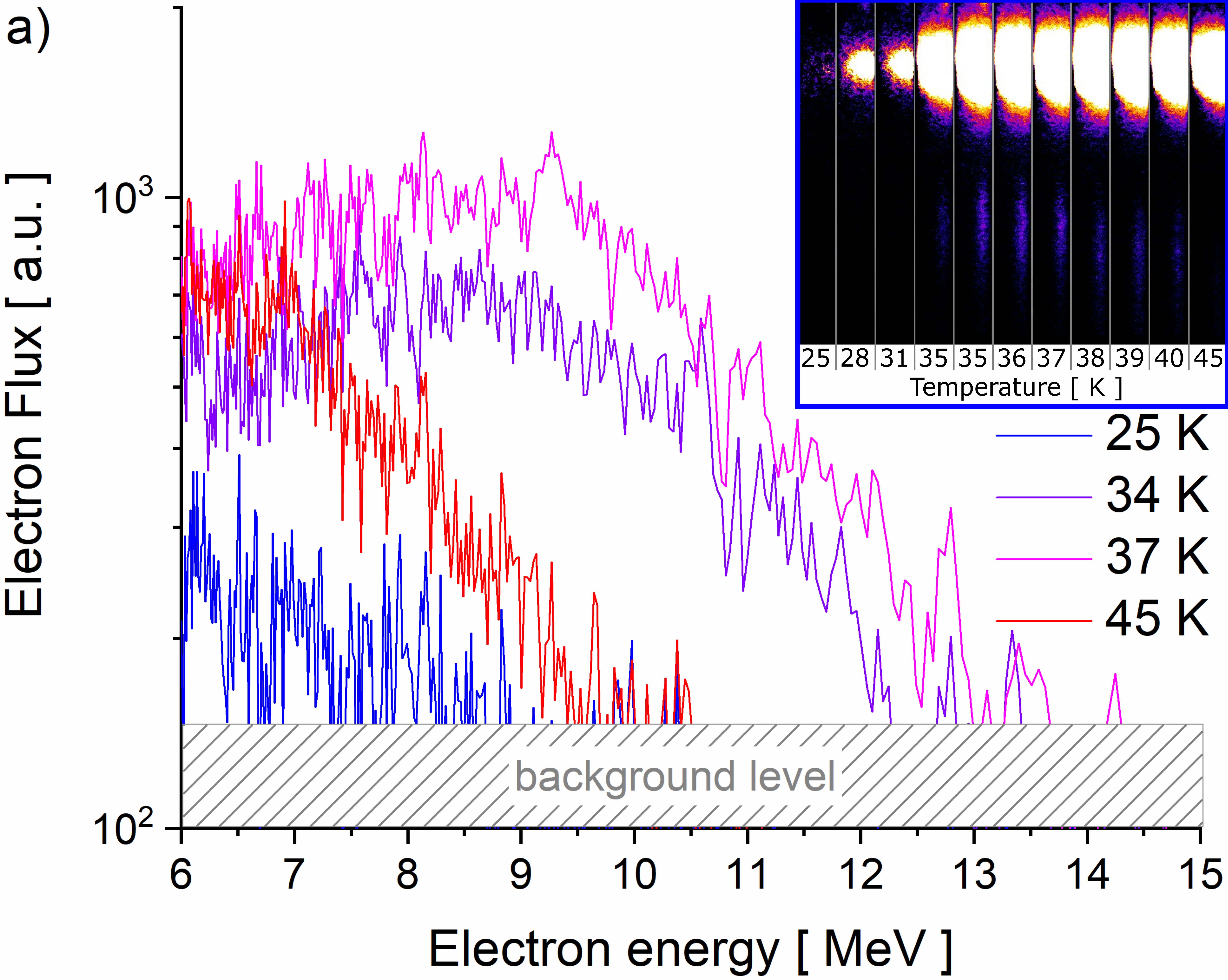}
	\includegraphics[width=\columnwidth]{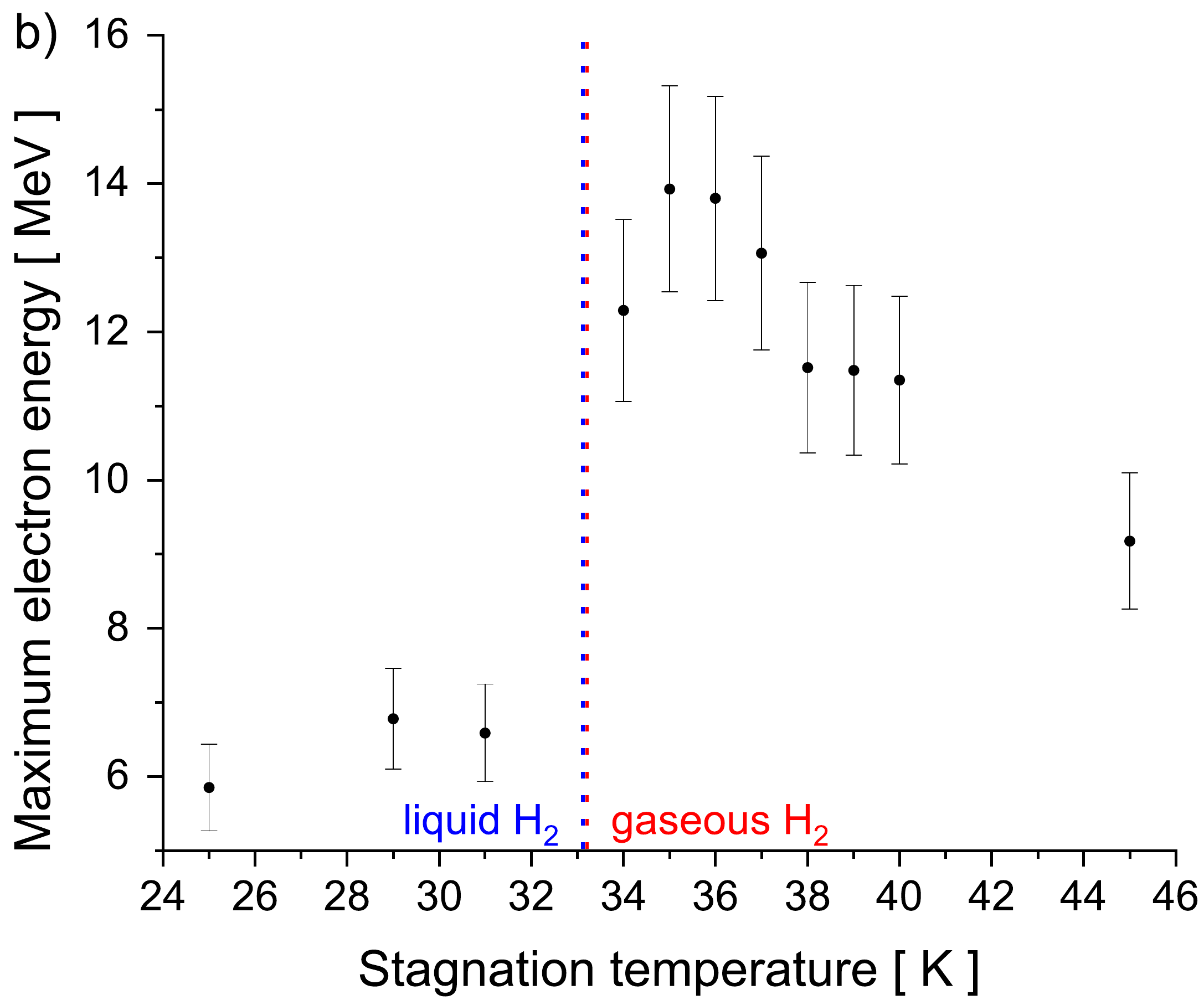}
	\caption{\label{fig2} a) Representative number of electron energy-spectra for different gas-temperatures (insert: raw data images). b) Evaluated maximum electron energy for the full temperature scan. The highest electron energy is achieved for the gaseous condition at the proximity between liquid and gaseous phase.}
\end{figure}

The electron energy in the laser forward direction was measured by means of the electron spectrometer to investigate electrons in the MeV energy range. The dependence of the maximum electron energy on the stagnation temperature for fixed laser parameters (E$_\mathrm{on target}$= 1.6\,J, $\tau$=30\,fs, p-polarised light) is shown for a representative number of spectra, each averaged over five shots. Some raw data images are shown in the insert of fig. \ref{fig2}a). Note that the signal obtained on the MCP detector to our finding was very weak and became only visible for a high MCP-amplification factor of $\approx$10$^5$. Even for this amplification the signal was close to the background noise level, which indicates a very low flux of these high-energetic electrons. The overall spectral shape of this high energy tail is a thermal distribution. The corresponding maximum  detected energy, above the background detection limit obtained for each temperature set, is presented in fig. \ref{fig2}b). This maximum energy is, for the case of liquid hydrogen in the reservoir forming the cluster jet (T $<$ 33.4\,K), significantly lower compared to the case in the proximity of the phase transition for which the hydrogen in the reservoir is already gaseous.
Right at the transition point the highest electron energy is obtained, a further increase in temperature led to a decrease in the maximum energy.

In the next step, the spatial profile of the electron signal, depending on laser energy and laser pulse duration, was investigated, using the IP-based setup presented in fig. \ref{fig1}b). Due to the necessity of warming up and cooling down the cryogenic-cluster target in order to replace the IP-stack, the parameter set was limited to two stagnation temperatures in the gaseous range,  (34\,K -- close to the transition temperature -- and 40\,K), for which the incidence laser light was varied between p- and c-polarisation.
For each of the four cases, defined by the parameter-set above, the dependence on the laser-energy and pulse-duration was investigated. Therefore one scan with three different settings -- either three different laser energies or three different pulse-duration -- in one pump-down was achieved, keeping the stagnation temperature and the laser polarisation constant.

For the energy scan, three consecutive shots with an on-target energy of $E_\mathrm{1}\approx$\,1.6\,J, $E_\mathrm{2}\approx$\,0.2\,J and $E_\mathrm{3}\approx$\,0.02\,J, while for the pulse-duration three consecutive shots with $\tau_\mathrm{1}\approx30$\,fs, $\tau_\mathrm{2}\approx\,300$\,fs and $\tau_\mathrm{3}\approx$1\,ps were recorded. Note that for the lowest laser energy ($\approx$0.02\,J) and the longest pulse-duration ($\approx$1\,ps) only very weak or no signal at all was detected.

\begin{figure}[ht] 
 	\centering	
 	\includegraphics[width=\columnwidth]{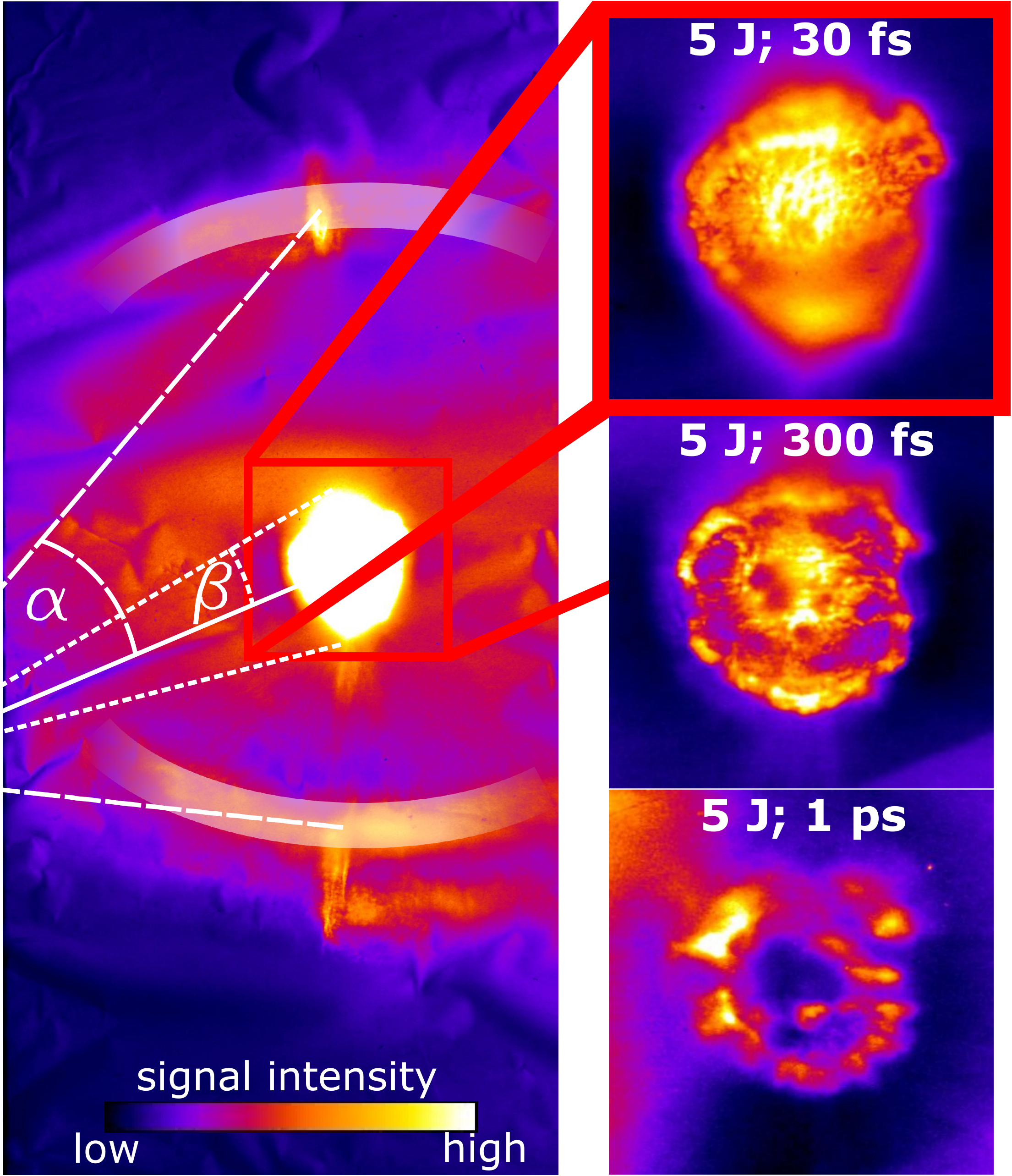}
	\caption{\label{fig3}Raw data image of the IP showing the signal in the laser forward direction depending on the laser pulse duration, covering an intensity range of two orders of magnitude, by varying the laser energy. The angles $\alpha, \beta$ denote the opening angles of the outer ring (highlighted in white) and the central spot, respectively.}
 \end{figure}

\begin{figure}[ht] 
 \centering	
 	\includegraphics[width=\columnwidth]{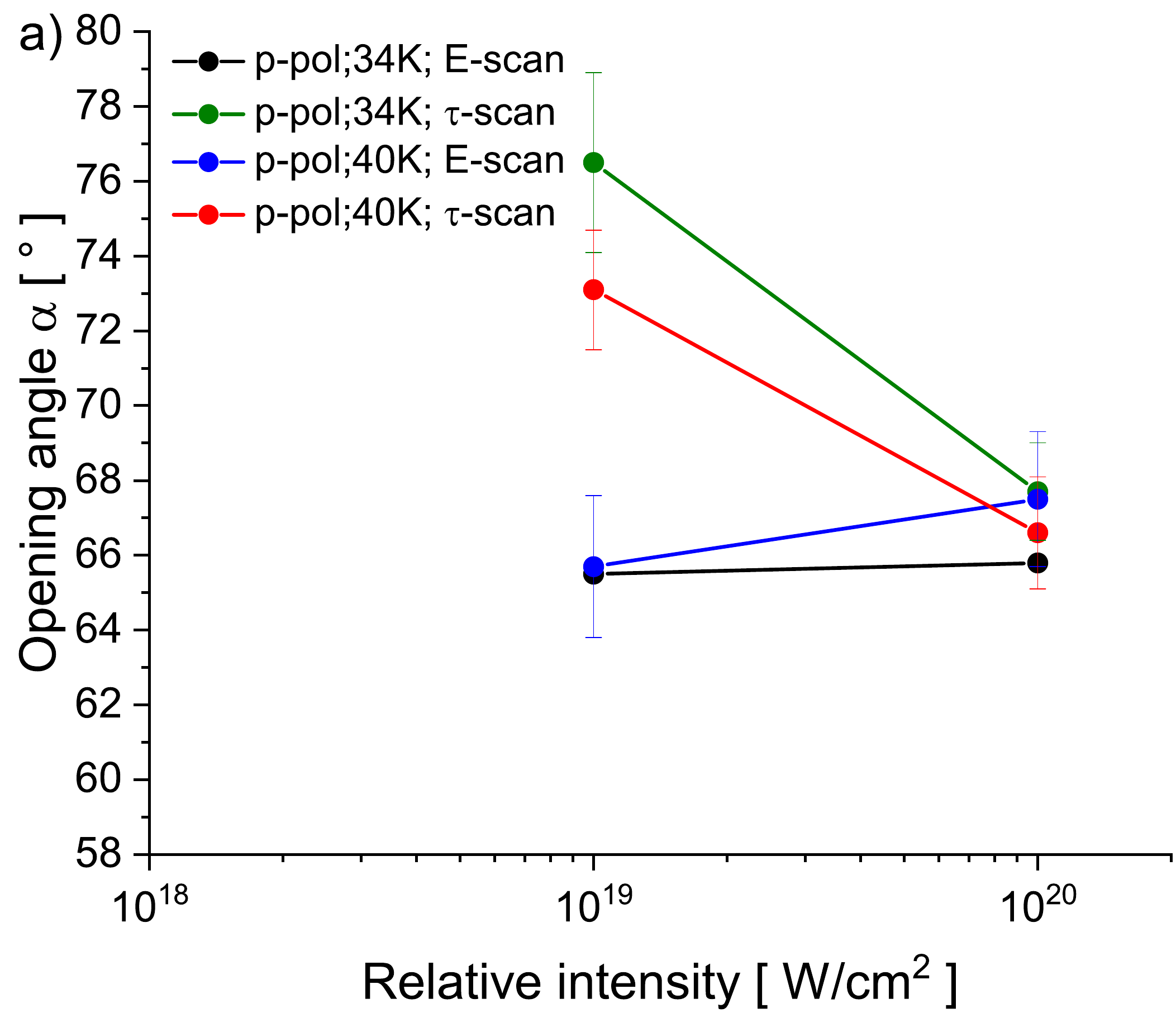}
	\includegraphics[width=\columnwidth]{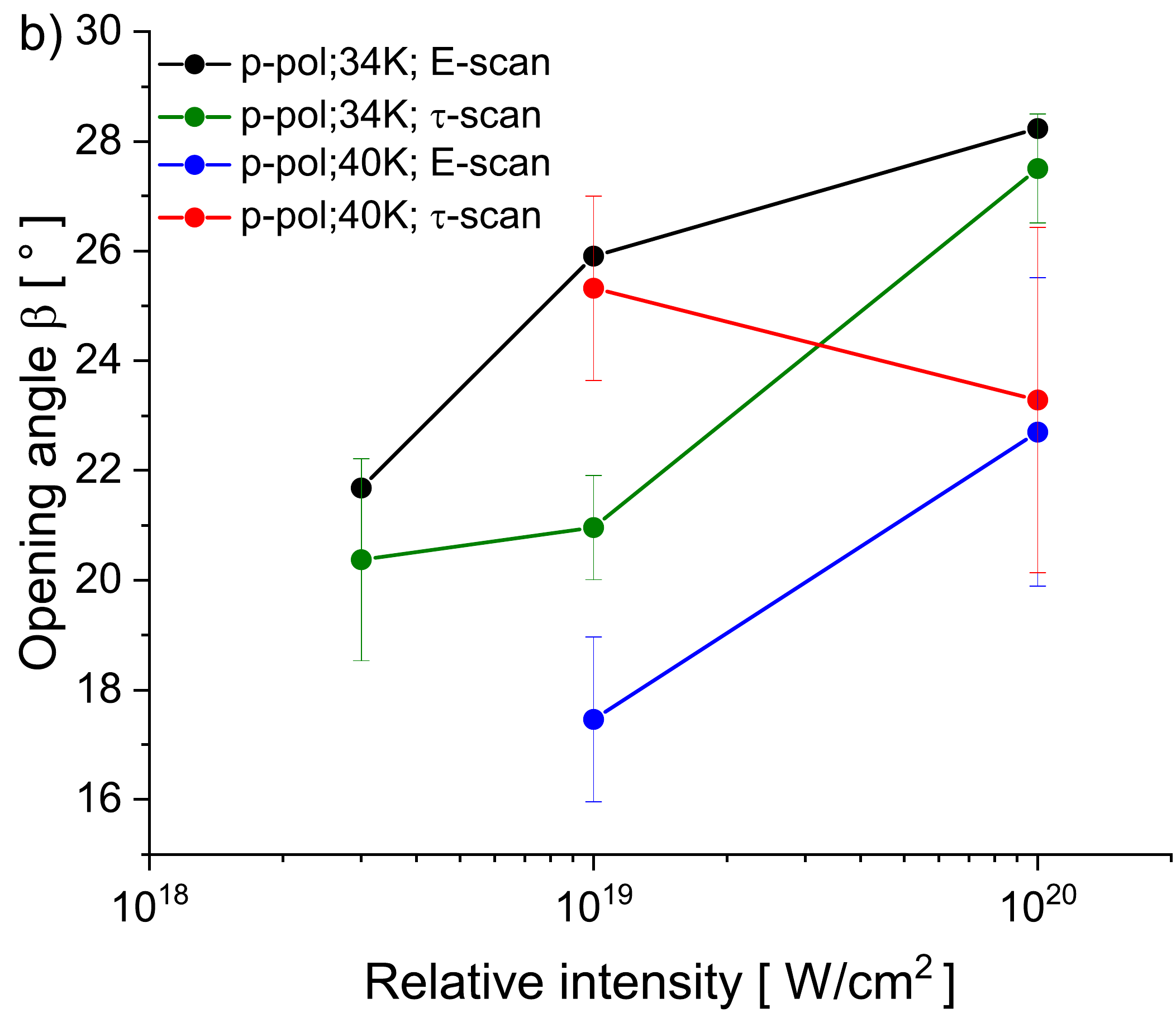}
 	\caption{\label{fig4}a) Opening angle $\alpha$ of the outer ring-structure and b) opening angle $\beta$ of the central spot, both depending on the laser- and/or target settings.}
 \end{figure}

As sketched out in fig. \ref{fig1}c), the main features occurring were an outer ring-structure and a central spot in the laser forward direction. This central spot contains areas which punctually saturates the read-out scanner within the first scan. Consecutive re-scanning  -- typically two or three times -- of the IP reduces the signal evenly, as described by Nave \textit{et al} \cite{Nave2011}, which allows to investigate the sub-structure of the central spot. As seen in fig. \ref{fig3} this spot has as well a ring-like structure observed within the filamented signal. The filamentation of the forward directed spot is caused by Weibel instabilities \cite{Weibel1959} due to thermal anisotropies in the plasmas, as e.g. observed by Quinn \textit{et al} \cite{Quinn2012} or Bulanov \textit{et al} \cite{Bulanov1998b}. The general decrease in signal intensity for increasing pulse length (fig. \ref{fig3}) might be attributed to a decrease in LWFA electrons: longer pulses change the shape of the electronic cavity leading to a different trapping condition. Generally, the laser pulse should perfectly fit in the first half of the plasma period to obtain efficient wakefield acceleration \cite{Pukhov2002}. Further, electrons accelerated via DLA or ponderomotive scattering can be ejected at different angles compared to the case of shorter pulse length. The opening angle $\alpha$ for the outer ring-structure and $\beta$ for the central spot were determined respectively. Therefore, three projections recorded from each IP signal were taken. Along each of these projections a multi-gauss fit was done and the distance between the peaks -- corresponding either to the ring-structure or to the central spot -- was taken to calculate the angles $\alpha$ and $\beta$ (see fig. \ref{fig1}c and \ref{fig3}). This method of gauss-fitting comes with the benefit of a representative result, even so the signal intensity in between different settings, especially the relation between the signal intensity of the outer ring structure and the central spot varied significantly, which is subject of a further investigation. The laser intensity is used as a harmonised scaling of the abscissa. Fig.\,\ref{fig4}a) shows the opening angle $\alpha$ for the outer ring-structure while fig.\,\ref{fig4}\,b) displays the opening angle $\beta$ of the central spot for the different scan parameters.

The influence of the geometry of both the outer ring structure and the central spot caused by a change of a factor of ten in the laser intensity is very small. Neither the opening angle $\alpha$ which is in between 60$^{\circ}$ and 80$^{\circ}$ nor the opening angle $\beta$ which is in between 17$^{\circ}$ and 28$^{\circ}$ exhibit a clear trend with respect to the laser intensity. For the process of ponderomotive scattering it is well known that particles with larger energy gain are ejected at a smaller angle than lower-energy ones \cite{Hartemann1995} which is in agreement with the intensity dependence of $\alpha$ for the data measured at 34 K. The increase of $\beta$ with the intensity hints at the fact that not only ponderomotive scattering might be responsible for the central spot structure. The different trends of the measurements taken at 34 K compared to those at 40 K might also be attributed to the fact that the lower temperature is closer to the target's phase transition. Similar opening angles $\alpha$, $\beta$ were obtained using a c-polarised laser pulse instead (data not included in fig. \ref{fig4}). Aside the stable geometrical aspect, a significant change in signal intensity of the central spot in between a stagnation temperature of 34\,K and 40\,K was observed, which is addressed in further investigations.

\begin{figure}[ht] 
 	\centering	
 	\includegraphics[width=\columnwidth]{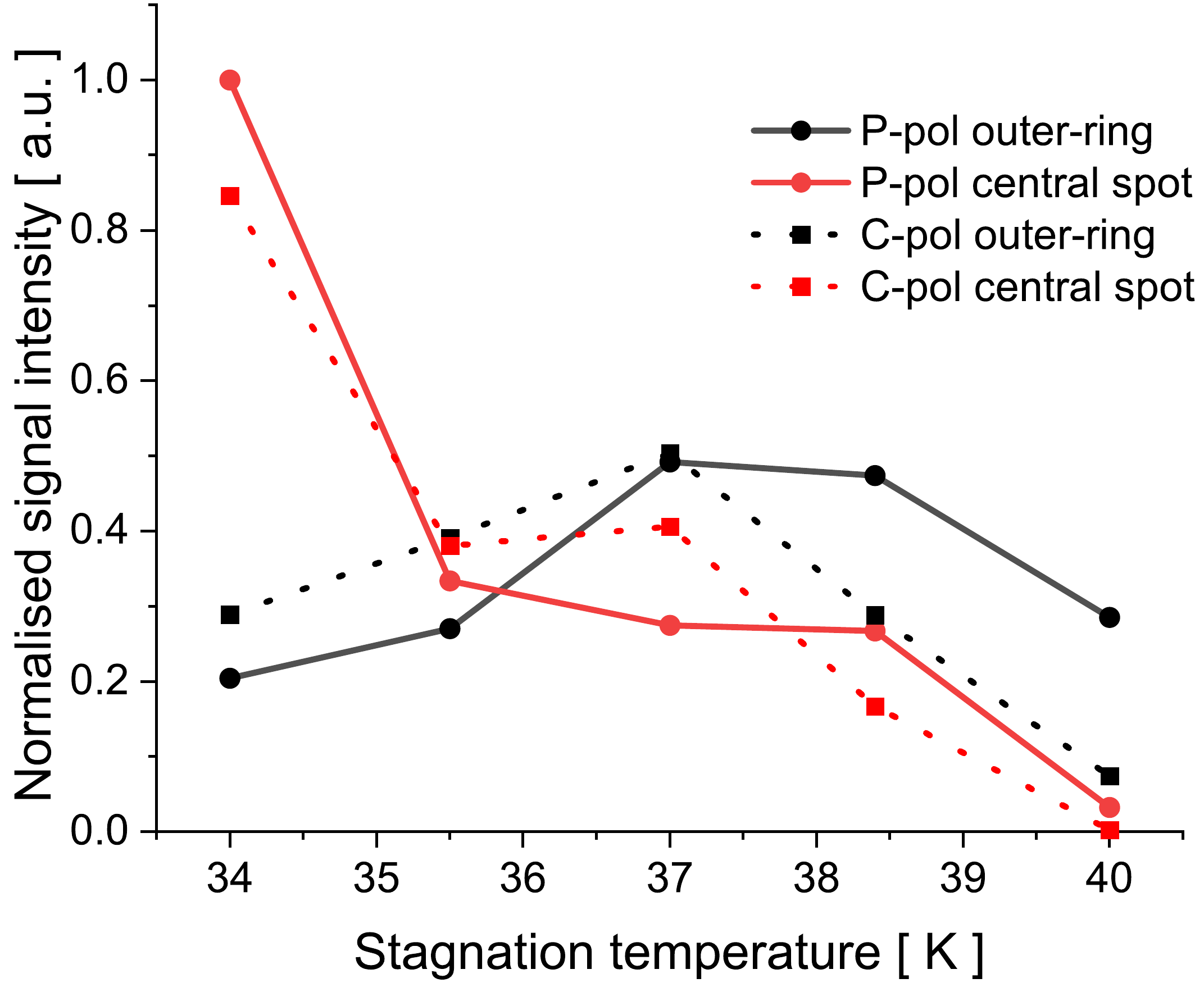}
 	\caption{\label{fig5} Temperature-dependent normalised signal intensity for the outer ring-structure and the central spot.}
 \end{figure}

Therefore another scan at the highest laser intensity (E$_\mathrm{target}\approx1.6$\,J; $\tau_\mathrm{target}\approx30$\,fs) was obtained while the stagnation temperature was varied in steps of 1.5 K for c- and p-polarised light. Fig. \ref{fig5} shows the normalised signal intensity for the outer ring-structure and the central spot. While the signal and thus the number of electrons of the outer ring-structure is more or less independent from the temperature, the central spot signal in laser forward direction decreases with increasing temperature. For a further understanding of the origin of these two clearly different electron patters, detailed 3D-PIC simulations were performed as discussed in the next section.
 
\begin{figure}[ht] 
  	\centering	
  	\includegraphics[width=\columnwidth]{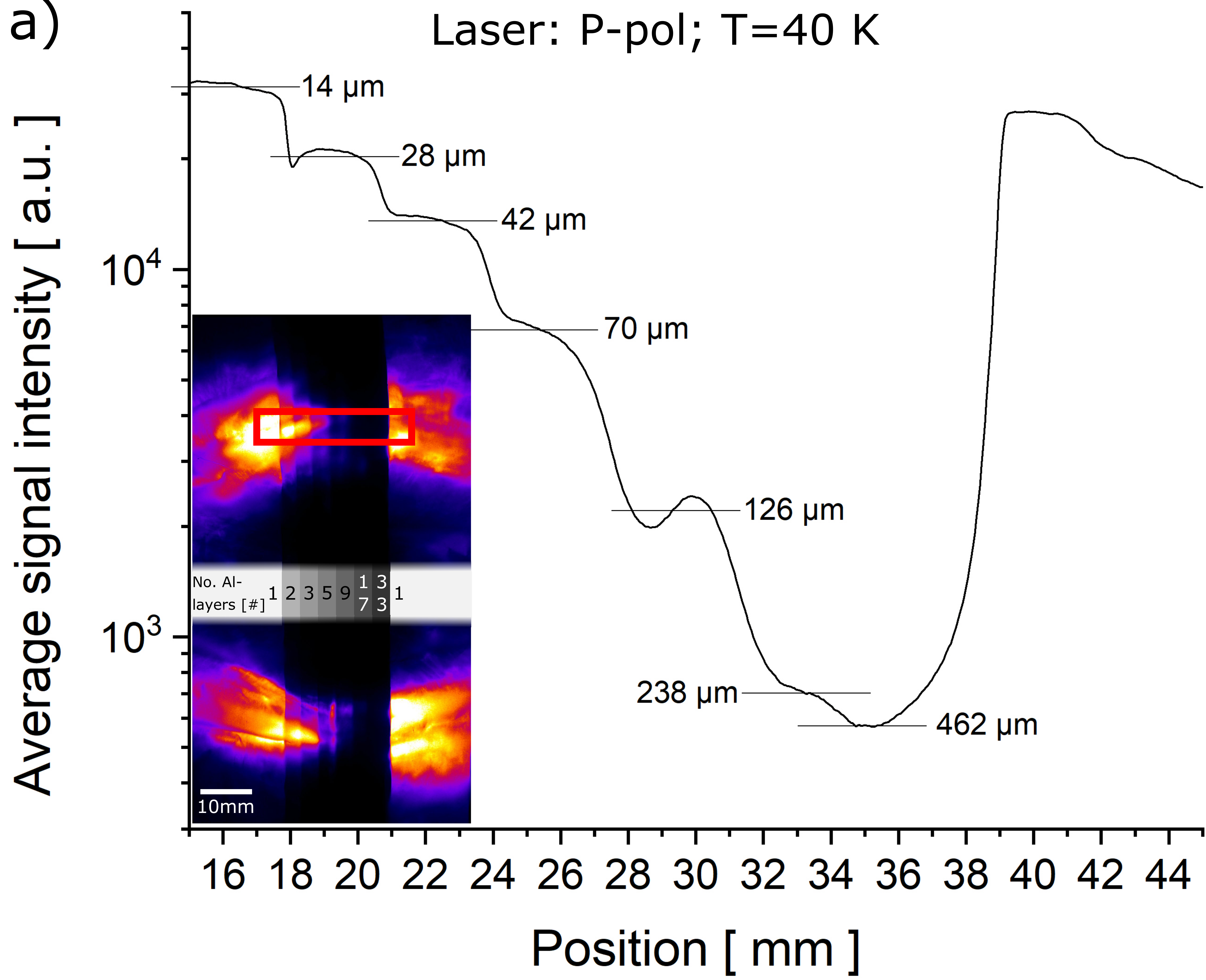}
  	\includegraphics[width=\columnwidth]{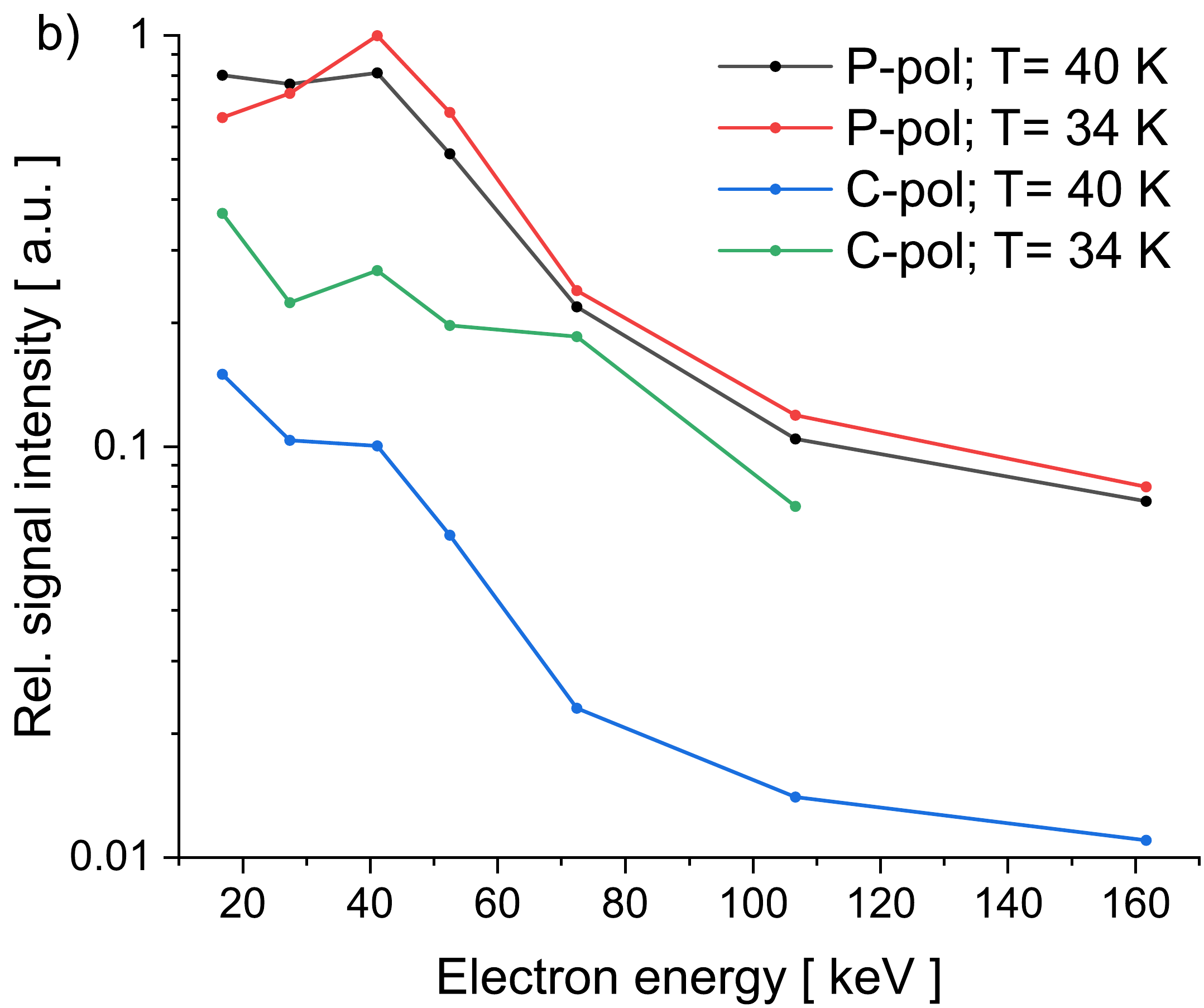}
  	\caption{\label{fig6} a) Signal line-out along the filter stack on the electron ring structure (insert: raw data image with filter). b) Evaluated relative signal intensity depending on the stopping energy for the different polarisation and temperatures obtained.}
  \end{figure}

  While the electron spectrometer in forward direction only covers a very small portion of high energetic electrons, a 14\,$\upmu$m thick aluminium foil was folded in an overlapping way to generate a filter stack with different thicknesses between 14 $\upmu$m and 462 $\upmu$m in front of the IP to estimate the electron energy of the outer ring structure. The signal was taken for p-polarised and c-polarised laser irradiation at stagnation temperatures of 34\,K and 40\,K. Fig. \ref{fig6}a) shows a typical line-out taken from IP-image, displayed as insert. For the different filter thicknesses the corresponding stopping power was calculated using the NIST \textit{EStar}-Database \cite{NIST2005}. Fig. \ref{fig6}b) gives the relative intensities for the different settings, background corrected and normalized to the highest occurring value. In the case of p-polarised light, the signal is stronger compared to c-polarised light, while at the same time slightly higher electron energies occur. Note, that the structure in laser forward direction got fully blocked by 2 or 3 layers of aluminium (not visible in fig.\,\ref{fig6}a)). This is in agreement with our finding that the electron signal detected with the spectrometer and the MCP was very weak or almost negligible, since the spectrometer cannot resolve electrons below 6\,MeV. At the same time, the few high energetic electrons are not able to generate a signal on the IP and become therefore not visible at higher filter thicknesses.
  
\section{Theory}\label{sec:PIC}
As already mentioned in the introduction, several acceleration processes are responsible for the structures visible on the IPs. To gain a more detailed understanding of the interaction between the laser pulse and  the clusters, numerical simulations using the PIC code \textsc{vlpl} \cite{Pukhov1999, Pukhov2016} were carried out.
The simulations use a box co-moving with the laser pulse with a size of ($100 \times 46 \times 46) \lambda_L^3$ and grid size $h_x = 0.05 \lambda_L$, $h_y = h_z = 0.1 \lambda_L$ ($x$ being the direction of laser propagation and $\lambda_L = 800$nm is the laser wavelength). The time step is chosen in accordance with the RIP solver \cite{Pukhov2020}, i.e. $\Delta t = h_x / c$.

For the first simulation, periodically placed clusters without any homogeneous background are considered. It is assumed that no prepulse is present, i.e. the clusters only interact with the peak pulse.
The clusters have radii of one cell size and are simulated with $10^4$ particles per cell for the electrons, while 500 particles per cell are used for the protons. The distance between two clusters is $3\lambda_L$ in any direction which is the distance estimated for clusters with 60 nm diameter, according to the parameters of the experiment. Thus, for reasons of computational effort, it is implicitly assumed that the clusters have expanded to some smaller extent and have an accordingly lower density of $n \approx 5 n_{crit}$. The interaction volume of periodically placed clusters is in total $375 \lambda_L$ long.
The parameters for the linearly polarised laser pulse are based on the experiment, i.e. $a_0 = 6.84$, $w_0 = 5$ $\upmu$m, $\tau = 30$ fs. The laser is assumed to exhibit a Gaussian profile and is focused towards the middle of the interaction volume.

\begin{figure*}
	\centering
	\includegraphics[width=\textwidth]{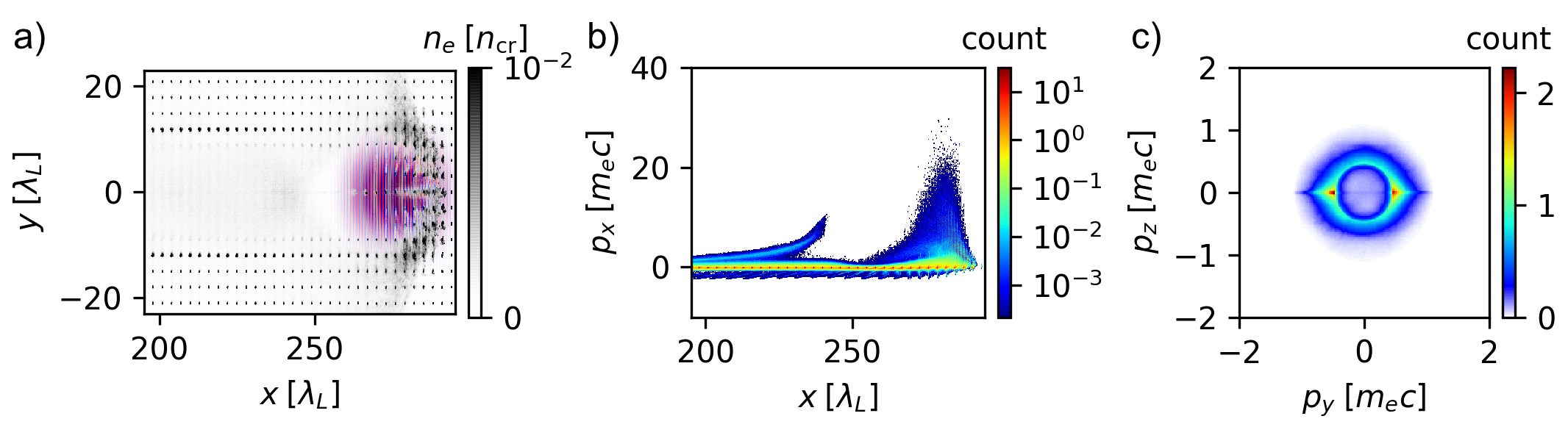}
	\caption{\label{fig7} PIC-simulation results for the laser pulse propagating through periodically placed clusters after $300T_0$. Shown in a) is the electron density overlayed with the $E_y$-field of the laser pulse. The electrons are mostly ponderomotively scattered in the transverse direction. In b) the longitudinal momentum of the electrons can be seen, while c) shows the $p_y$-$p_z$ phase space.}
\end{figure*}

As can be seen in fig. \ref{fig7}a), the laser pulse rips electrons from the clusters and expels them in the transverse direction. This creates a structure reminiscent of the cavity in wakefield acceleration, although only a few electrons are wakefield-accelerated (see region $x < 250 \lambda_L$ in \ref{fig7}b)). The acceleration gradient $E_x$ in the absence of any initial background is very weak, thus no high-energy electrons are to be expected from this process here.
Looking at the phase space of the electrons (fig.\ref{fig7}b) and \ref{fig7}c)), it can be observed that most of the electrons are directly pushed by the laser pulse. There are two processes prominent for accelerating electrons at the laser position, the first being DLA. For efficient acceleration with this mechanism it is necessary for the electrons to stay in phase with the accelerating parts of the laser field for as long as possible. A discussion of parameters relevant for DLA can be found in \cite{Jirka2020}. The acceleration of electrons via DLA over a significant period of time can be ruled out in the presented results, as the final electron energies would need to be orders of magnitude larger. It can also directly be seen in the PIC results that the peak momentum of the electrons shown in fig. \ref{fig7}b) is not increasing, i.e. the DLA electrons quickly dephase.

\begin{figure*}[t]
	\centering
	\includegraphics[width=\columnwidth]{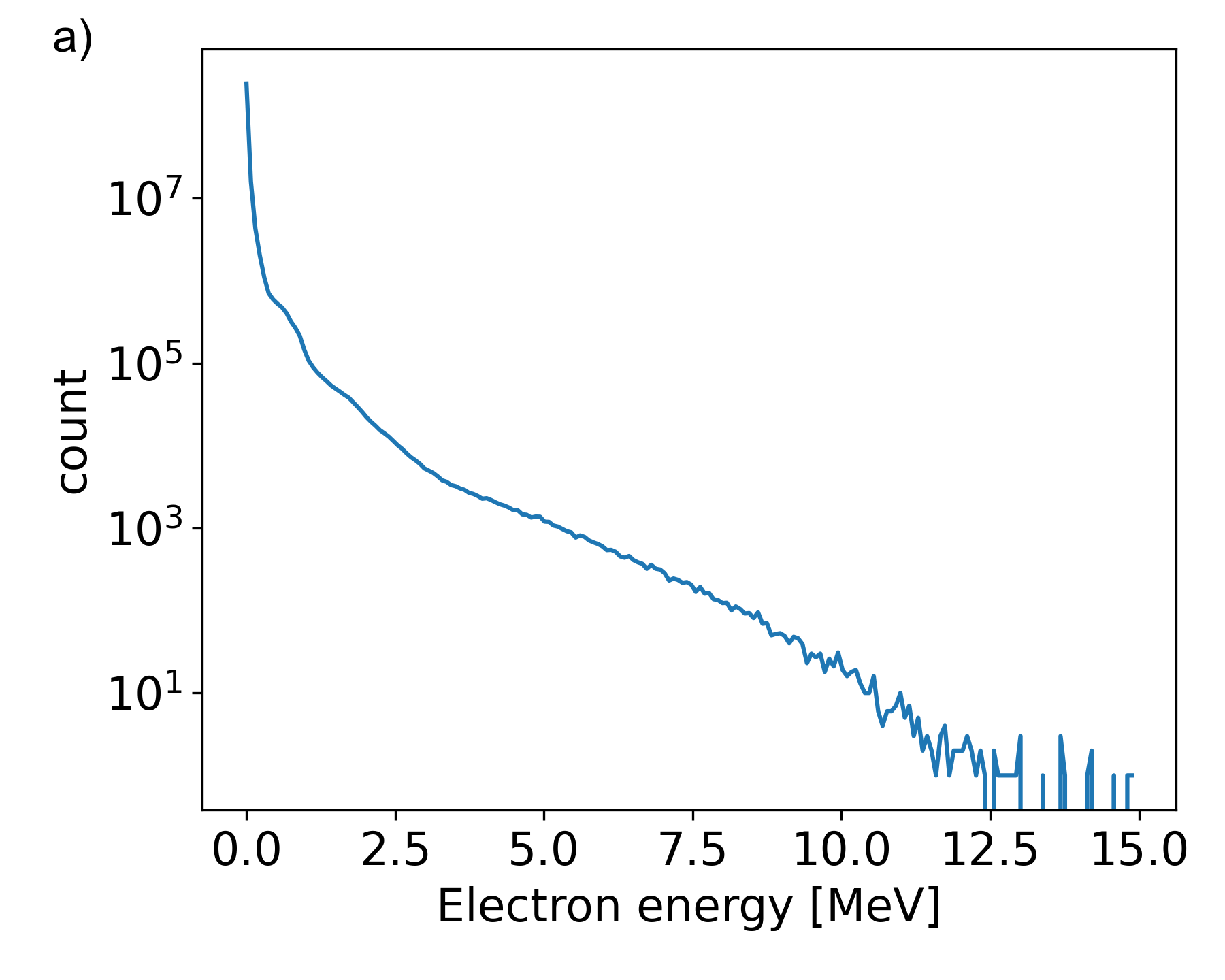}
	\includegraphics[width=\columnwidth]{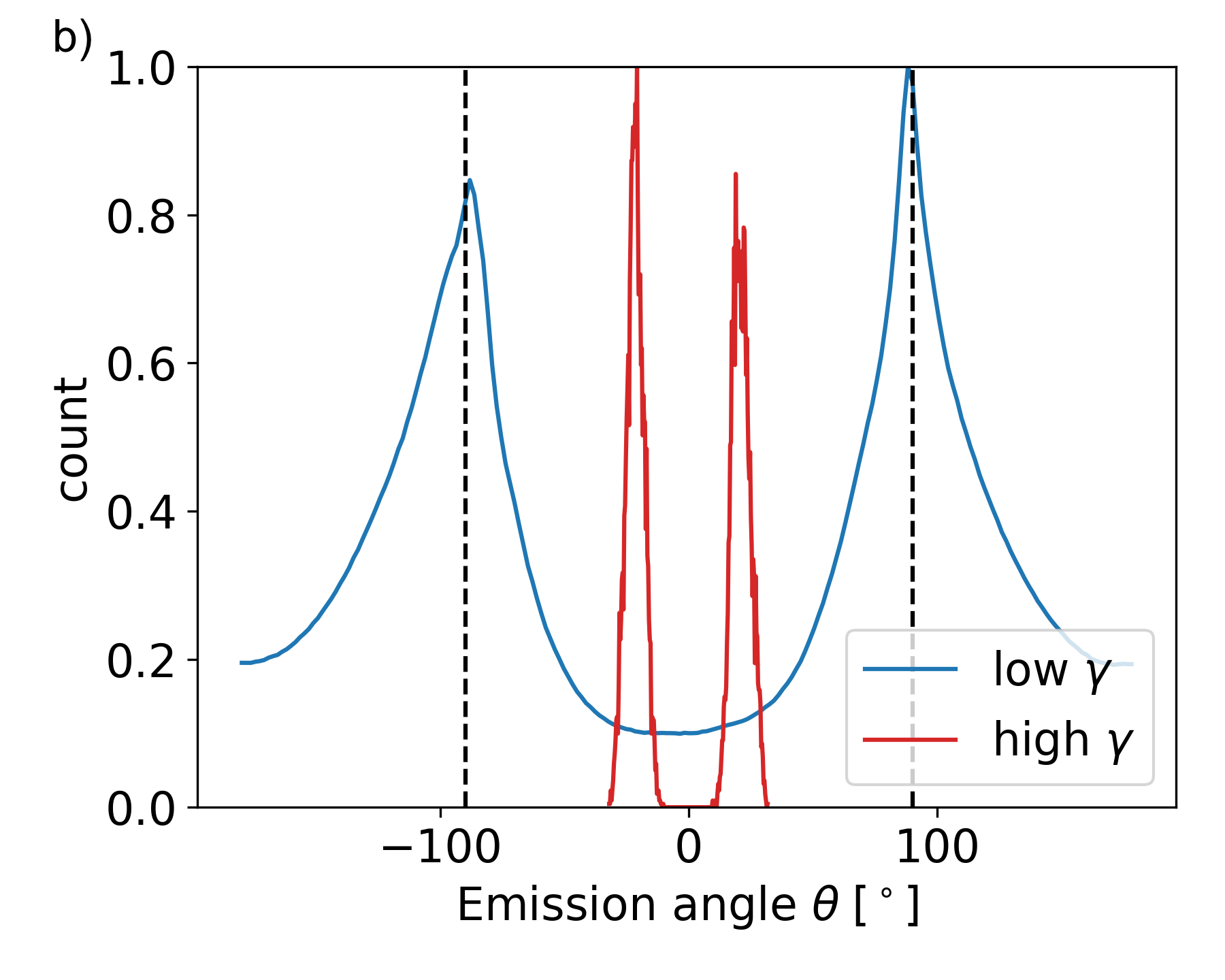}
	\caption{\label{fig8}a) Logarithmic energy spectrum of the electrons in the simulation box after $300T_0$. The spectrum is linearly decreasing and maximum kinetic energies of around 12 MeV can be observed. b) Angular spectrum for electrons at 300$T_0$ with lower (blue) and higher (red) energy (both normalised to 1 for better visibility). The corresponding energy range can be found in the text. The dashed lines denote angles of $\pm 90^\circ$.}
\end{figure*}

\begin{figure}
	\centering
	\includegraphics[width=\columnwidth]{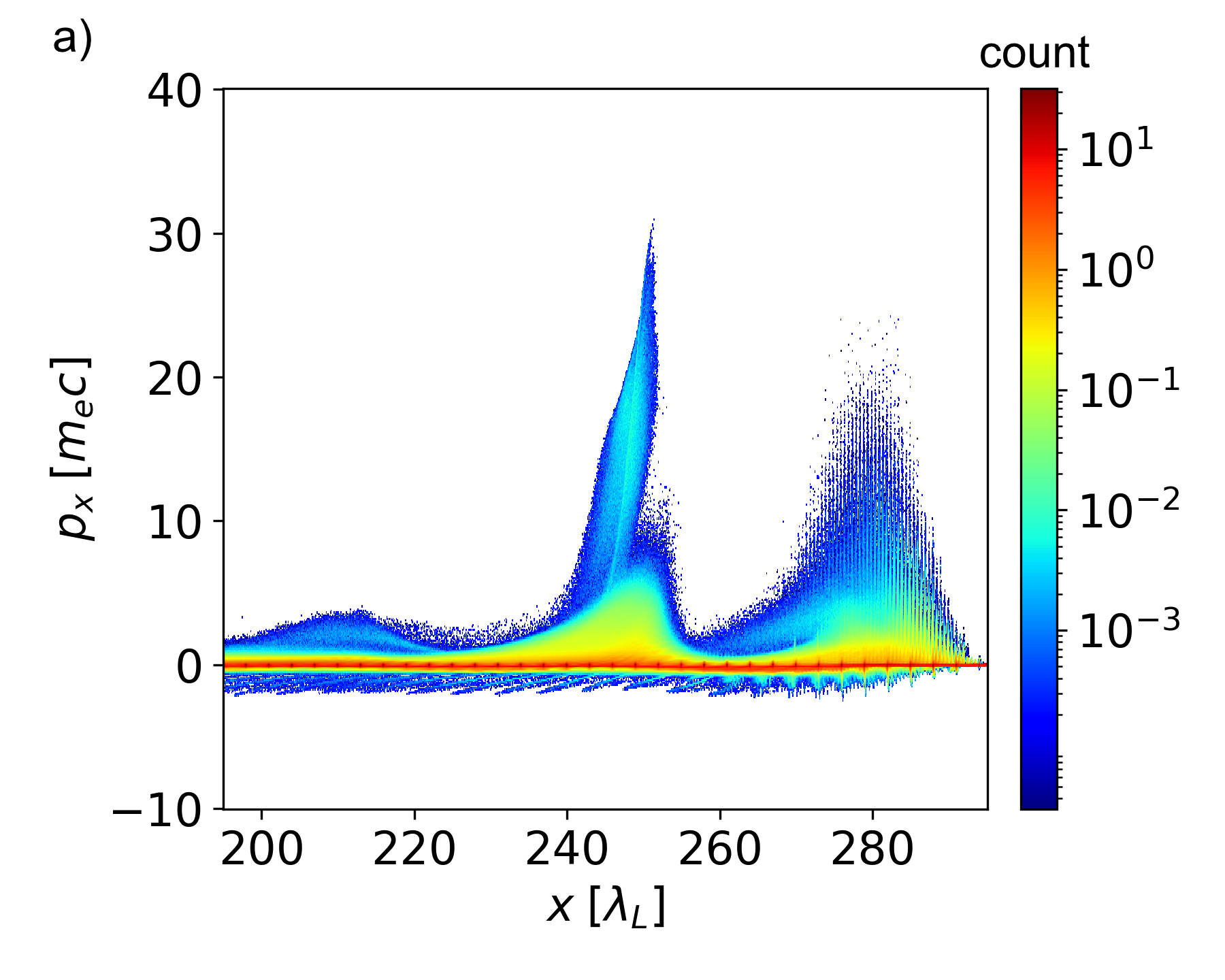}
	\includegraphics[width=\columnwidth]{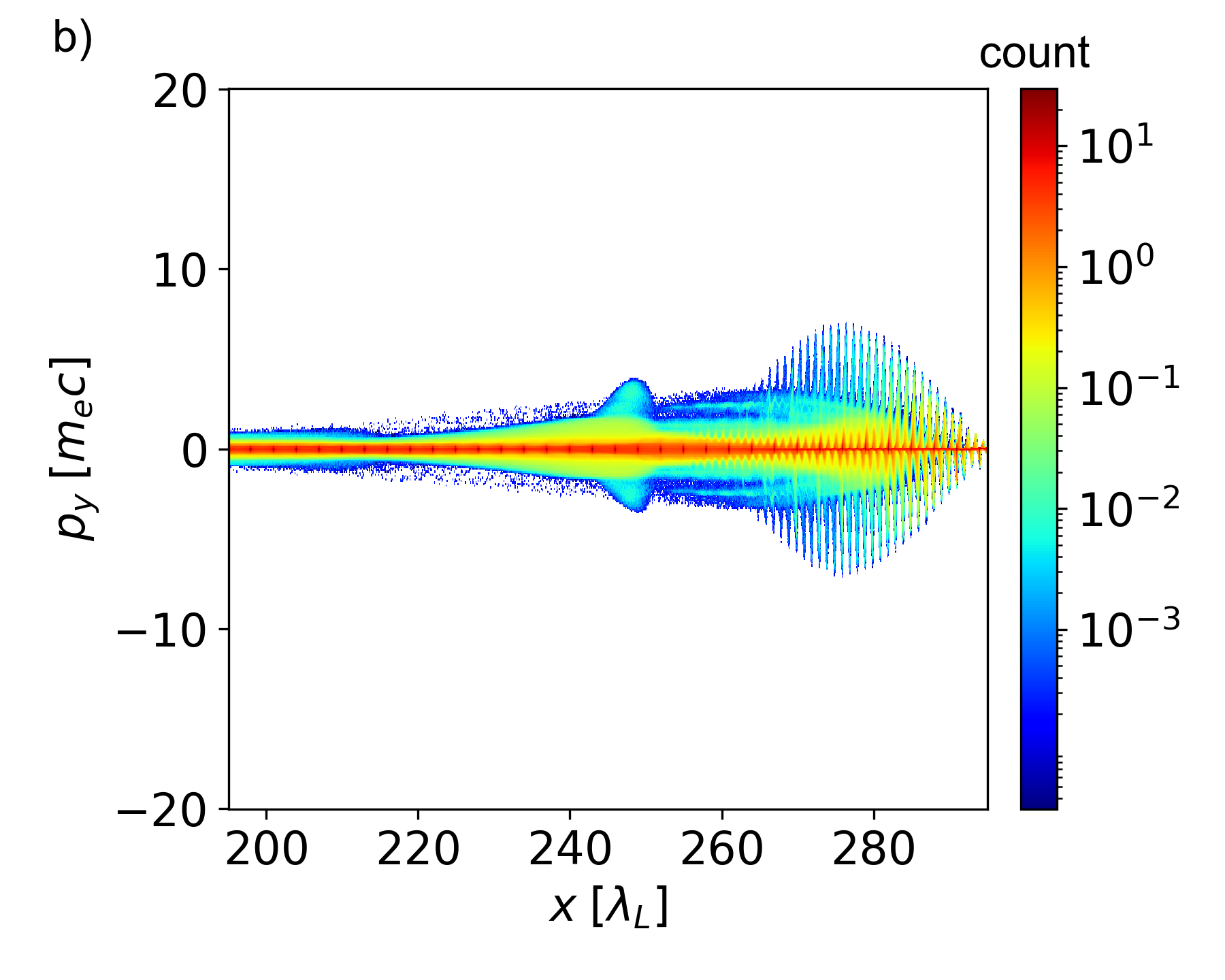}
	\caption{\label{fig9}Phase space plots for PIC simulations with a background density of $n_e = 5 \times 10^{17} \mathrm{cm}^{-3}$, a) showing the $x$-$p_x$-plane and b) the $x$-$p_y$-plane. As visible here, there are two electron populations: the one in the front stems from DLA and ponderomotive scattering and is visible also for no background present. The population in the back is accelerated by the gradient in $E_x$ which stems from the cavity.}
\end{figure}

The second process to be considered for acceleration is the ponderomotive force of the laser pulse directly pushing the electrons outwards. This has been analytically considered by \cite{Hartemann1995, Quesnel1998} and experimentally observed for energies in the range of a few keV by Monot \textit{et al} \cite{Monot1993} and about 100 keV by Moore \textit{et al} \cite{Moore1995}. The same process was also used as an explanation for electron acceleration in the MeV range by Malka \textit{et al} \cite{Malka1997}. Considering an electron that is initially at rest, i.e. the electron's Lorentz factor is $\gamma_0 \approx 1$, the emission angle $\theta$ caused by the ponderomotive force is given as
\begin{align}
	\tan(\theta) = \frac{\sqrt{2(\gamma / \gamma_0 -1) / (1 + \beta_0)}}{\gamma - \gamma_0 (1 - \beta_0)} \; , \label{eq:theta}
\end{align}
where $\beta_0 = v_0 / c$ is the electron's velocity prior to acceleration normalized to the speed of light \cite{Hartemann1995}. Since the electrons have not been pre-accelerated it is reasonable to assume that $\beta_0 \approx 0$ and therefore $\gamma_0 \approx 1$. Thus, eq. (\ref{eq:theta}) simplifies to $\tan(\theta) = \sqrt{2 / (\gamma  - 1)}$. This equation further assumes that the electron is initially located on the central axis ($y = 0$) and that it experiences the maximum electric field in the laser focus position.

Using the experimentally obtained values for the opening angles of the outer ring (ca. $60^\circ$-$80^\circ$), this equation yields electron energies in the range of 30 keV up to 300 keV. The opening angles for the central spot were in the range of $16^\circ$ to $30^\circ$, corresponding to energies in the range of roughly (3-12) MeV. 
Therefore, the analytical approximation fits the experimental data rather well. The deviations are likely to be explained by the fact that not all electrons are ejected from the central axis and not all electrons stems from the location of laser focus. Further, the equation generally only holds for acceleration in vacuum. The presence of any background plasma and clusters will modulate the resulting energy spectrum.
Lastly, deviations of the of the nominal peak intensity and focal spot size of the laser in experiment could explain some of the differences as lower laser power would decrease the emission angle.

Hartemann \textit{et al}, in their paper \cite{Hartemann1995} also give an approximation for the maximum electron energy that can be achieved using ponderomotive scattering, namely
\begin{align}
	\gamma_\mathrm{max} = \gamma_0 \left[ 1 + \frac{a_0^2(1 + \beta_0)}{2}\right] \; .
\end{align}

Again, assuming that $\beta_0 \approx 0, \gamma_0 \approx 1$ yields $\gamma_\mathrm{max} \approx 24.4$ for a laser pulse with $a_0 = 6.84$ like it is used in the experiment. This is in good agreement with both the experimental data and the PIC simulations (cmp. experimental spectra in fig. \ref{fig2}a) and the PIC spectrum in fig. \ref{fig8}a)) which shows the energy spectrum of all electrons inside the simulation box after 300$T_0$.

For the PIC simulations, the emission angles corresponding to the electron energies can directly be compared: as to be expected, the ponderomotive scattering is radially symmetric. This can be seen rather well in fig. \ref{fig7}c) where the $p_y$-$p_z$-plane for electrons with $1.1 \leq \gamma \leq 1.5$ is shown. The additional feature along the line $p_z = 0$ comes from the fact that a linearly polarised laser pulse is used. For circularly polarised pulses this feature vanishes leaving behind a fully symmetric ring (not shown here).
In figure \ref{fig8}b) the emission angles at 300$T_0$ for electrons with kinetic energies $20 \; \mathrm{keV} \leq E \leq 160 \; \mathrm{keV}$ (blue line) and electrons with kinetic energies $6 \; \mathrm{MeV} \leq E \leq 12 \; \mathrm{MeV}$ (red line) are plotted as histograms. Here, we set $\tan (\theta) = p_y / p_x$, thus leaving out the aforementioned asymmetry in $z$-direction due to the polarisation plane. The lower-energy electrons clearly make up a ring structure with main emission angles of slightly below $90^\circ$, which is in good correspondence with the experimental and analytical values. The emission angle for the central spot also fits well with angles in the range of $20^\circ$-$30^\circ$. It has to be noted, that both lines are independently normalised to a maximum value of 1 for better visibility, i.e. the signal in forward direction is weaker than the signal of lower energy electrons as visible from the full energy spectrum. Again, some deviations compared to the experiment might stem from the fact that the nominal laser power is not quite reached.

While the data presented for the case of clusters interacting with only the laser peak pulse fit rather well, it needs to be stressed that this is not the complete physical picture. The prepulse will already rip off electrons from clusters and lead partly to Coulomb explosion. Thus, when the peak pulse reaches the point of interaction after some ps, parts of the cluster target will already have homogenised. The importance of the prepulse for clustered targets was already considered by Auguste \textit{et al} in \cite{Auguste2000}.
If, for comparison, it is assumed that the peak pulse only interacts with a completely homogeneous underdense electron target, ponderomotive scattering is still mainly responsible for the formation of the outer ring (not shown here). Wakefield acceleration for densities in the region $10^{16}-10^{17}$ cm$^{-3}$ is largely negligible, although it has to be noted that due to the large box size only a smaller number of particles per cell can be used compared to the case of the clustered target. Only some electrons at the wake's tail get weakly accelerated.
Further, the absence of any clusters also reduces the maximum energy of the electrons observed in the simulations.

The peak pulse therefore likely sees an intermediate target: While many of the clusters already have been destroyed by the prepulse, some remainders of clusters are still present, leading to a homogeneous electron target modulated with some clusters.
Taking the simulation of the purely clustered target from before and adding a homogeneous background of $n_e = 5 \times 10^{17} \mathrm{cm}^{-3}$ strongly overestimates the density of the experimental target but shows the changes of this transition regime compared to the cases of only clusters and purely homogeneous background more clearly.
More particles are introduced at the wake's stern (cmp. fig. \ref{fig7}b) and \ref{fig9}a)). These electrons are located towards the end of the cavity created by the laser pulse ($x < 260 \lambda_L$) and contribute to the spatial distribution of the accelerated electrons. Increasing the background density further would lead to a stronger $E_x$ accelerating those electrons. Moreover, the stronger transverse fields for higher background densities would reduce the spread in $p_y$ seen in fig. \ref{fig9}b), i.e. an increase in density would focus more electrons into the central spot seen on the IPs. A discussion of the influence of clusters for LWFA is given by Mayr \textit{et al} \cite{Mayr2020}. The presence of clusters introduces highly localised changes in potential that can give electrons the momentum kick needed to be injected into the wake. Depending on the charge density and the diameter of those clusters, the electrons can be more easily injected into the wakefield. This method of injection is similar to the technique of ionisation injection \cite{Oz2007}, where different (heavier) particle species are used to inject electrons due to the local changes in potential.

While the simulations already reproduce the experimentally observed energy spectra and emission angles quite well, the results could further be optimised by randomising the particles' positions and introducing clusters of various sizes into the interaction volume. As seen by Mayr \textit{et al}, randomisation of cluster location leads to some changes in the results while leaving the main features of the acceleration scheme intact \cite{Mayr2020}.

\section{Conclusion}
We demonstrated that cryogenic hydrogen cluster-targets allow for quickly replenishable sources of hundred-keV electrons using a high-intensity laser via ponderomotive scattering.
The influence of laser-parameters like intensity and pulse length, as well as target parameters like the stagnation temperature and backing gas pressure, onto the resulting structure were examined.
Throughout the measurements, an outer ring structure could be observed that is largely robust against variation of parameters. Further, for lower temperatures, a central spot inside the ring structure could be seen.
It could be shown by PIC simulations that these structures stem from several acceleration processes although ponderomotive scattering is the most prominent. 
Further, the effect of the prepulse onto the target shape has been discussed. It homogenises the clusters partly before the peak pulse hits creating a transition regime between the purely clustered and purely homogeneous case.
Depending on the background density and the cluster parameters, the wakefield acceleration and direct laser acceleration can gain importance for the formation process of the electronic structure.
These outer electrons can be used i.a. for the generation of transition radiation in the THz range by interaction with a foil \cite{Yang2018} as well as lower-energy, but nC-class sources \cite{Brunetti2017} needed for e.g. radiotherapy.

\begin{acknowledgments}
We acknowledge financial support by the German Federal Ministry of Education and Research (joint research project; No. 05K2016). The theoretical study was in part supported by the DFG (project PU 213/6-1).
The authors gratefully acknowledge the Gauss Centre for Supercomputing e.V. (www.gauss-centre.eu) for funding this project (qed20) by providing computing time on the GCS Supercomputer JUWELS at J\"{u}lich Supercomputing Centre (JSC). L.R. would like to thank X.F. Shen and K. Jiang for the helpful discussions throughout the the project.
\end{acknowledgments}

\bibliographystyle{unsrt}

\end{document}